\documentclass{iopart}

\usepackage{iopams}
\usepackage{graphicx}  
\usepackage{url}

\def\be{\begin{equation}}
\def\ee{\end{equation}}
\def\bea{\begin{eqnarray}}
\def\eea{\end{eqnarray}}

\begin{document}

\title[The second round of Mock LISA Data Challenges]{An overview of the second round of the Mock LISA Data Challenges}

\author{K A Arnaud$^1$,
S Babak$^2$,
J G Baker$^1$,
M J Benacquista$^3$
N J Cornish$^4$,
C Cutler$^5$,
L S Finn$^6$,
S L Larson$^7$,
T Littenberg$^4$,
E K Porter$^{2,4}$,
M Vallisneri$^5$,
A Vecchio$^{8,9}$,
J-Y Vinet$^{10}$ (The Mock LISA Data Challenge Task Force)}

\address{$^1$ Gravitational Astrophysics Laboratory, NASA Goddard Space Flight Center, 8800 Greenbelt Road, Greenbelt, MD 20771, US}
\address{$^2$ Max-Planck-Institut f\"{u}r Gravitationsphysik (Albert-Einstein-Institut), Am M\"{u}hlenberg 1, D-14476 Golm bei Potsdam, Germany}
\address{$^3$ Center for Gravitational Wave Astronomy, University of Texas at Brownsville, Brownsville, TX 78520, US}
\address{$^4$ Dept.\ of Phys., Montana State University, Bozeman, MT 59717, US}
\address{$^{5}$ Jet Propulsion Laboratory, California Institute of Technology, Pasadena, CA 91109, US}
\address{$^{6}$ Center for Gravitational Wave Physics, The Pennsylvania State University, University Park, PA 16802, US}
\address{$^{7}$ Dept.\ of Phys., Weber State University, 2508 University of Circle, Ogden, UT 84408, US}
\address{$^{8}$ School of Phys.\ and Astron., University of Birmingham, Edgbaston, Birmingham B152TT, UK}
\address{$^{9}$ Dept.\ of Phys.\ and Astron., Northwestern University, Evanston, IL 60208, US}
\address{$^{10}$ D\'epartement ARTEMIS, Observatoire de la C\^{o}te d'Azur, BP 429, 06304 Nice, France}
\ead{lisatools-mldc@gravity.psu.edu}
\begin{abstract}
The Mock Data Challenges (MLDCs) have the dual purpose of fostering the development of LISA data-analysis tools and capabilities and of demonstrating the technical readiness already achieved by the gravitational-wave community in distilling a rich science payoff from the LISA data. The first round of MLDCs has just been completed and the second-round data sets are being released shortly after this workshop. The second-round data sets contain radiation from an entire Galactic population of stellar-mass binary systems, from massive--black-hole binaries, and from extreme--mass-ratio inspirals. These data sets are designed to capture much of the complexity that is expected in the actual LISA data, and should provide a fairly realistic setting to test advanced data-analysis techniques, and in particular the global aspect of the analysis. Here we describe the second round of MLDCs and provide details about its implementation.
\end{abstract}


\section{Introduction}
\label{s:intro}

The Laser Interferometer Space Antenna (LISA) is a spaceborne gravitational-wave (GW) laser interferometer for the observation of the low-frequency  ($\approx$ 0.1 mHz--1 Hz) GW sky (see \cite{ScienceCase,lisappa} and Danzmann's contribution in this volume).  LISA is an all-sky monitor with the capability of observing a variety of compact-object binary systems, with masses ranging from a fraction to millions of solar masses.  Moreover, LISA could discover GWs from entirely new classes of sources, such as exotic compact objects and relics from the early universe (see \cite{ScienceCase,CT2002} and references therein).  Although much relevant experience has already been gained in the analysis of GW data collected by ground-based detectors, the differences between space-based and ground-based observations present new difficulties and require novel solutions for data analysis. These differences include the complex LISA response, the confusion noise from Galactic and extra-Galactic binary populations, and the simultaneous presence of many weak and strong GW signals. It is important to tackle these new analysis problems early, in order to develop the tools and methods necessary for the maximum science exploitation of such a revolutionary data set.

The LISA International Science Team (LIST) has embarked on a programme to foster the development and evaluate the technical readiness of data-analysis tools and capabilities for LISA. This programme goes under the name of Mock LISA Data Challenges (MLDCs). The MLDC Task Force\footnote{Full details regarding the Task Force activities, as well as links to relevant resources, are available on the Task Force wiki~\cite{MLDCwiki}.} has been charged by the LIST to formulate challenge problems, develop standard models of the LISA spacecraft and orbits, and of GW sources, provide computing tools (e.g., LISA response simulators and source-waveform generators), establish criteria for the evaluation of the responses to the challenges, and provide any technical support necessary to the challenge participants. These challenges are meant to be blind tests, but not really contests; the greatest scientific benefit stemming from them will come from the quantitative comparison of results, analysis methods and implementations.

The first round of MLDCs~\cite{MLDCLISA06a,MLDCLISA06b} has just been completed. Details on the data sets, techniques developed for the analyses, and results are provided in the companion article in this volume~\cite{MLDC1-gwdaw}, and in the references cited there. In this short paper we describe the second round of MLDCs, which has just been released~\cite{MLDCweb}. The Challenge-2 data sets represent a very significant increase in complexity with respect to those distributed for Challenge 1. More importantly, they contain the full set of key sources that are expected to make up the real LISA data, and therefore provide a realistic testbed for different data-analysis approaches. Since the instrument response, instrument noise, and waveform models adopted so far in the MLDCs employ various simplifications, much work will be needed beyond the completion of Challenge 2 to fully develop the data-analysis capabilities necessary for the mission, and future rounds of MLDCs will address progressively more realistic measurement scenarios and introduce more general gravitational waveforms. The details of the schedule of MLDCs from Challenge 3 onwards have not been decided yet, but we expect to release a third round of MLDCs in summer 2007.

\section{Mock LISA Data Challenges: Second round}
\label{s:MLDC2}

The Challenge-1 data sets contained single sources, or sets of several ($\approx 20$) signals well separated in parameter space; an exception were two data sets containing $\approx 50$ signals from Galactic binaries, concentrated in frequency bands of width 30 $\mu$Hz and 3 $\mu$Hz. Challenge 1 focused on only two signal classes: Galactic stellar-mass binaries and massive--black-hole (MBH) binary inspirals. Indeed, the key goal of this first round of challenges was the development and validation of source-specific data-analysis techniques for the sources featured in the LISA mission's minimum science requirements. Challenge 2 has a much more realistic flavour: it includes millions of Galactic binaries,
as we expect in reality, in addition to multiple massive--black-hole binaries and extreme--mass-ratio inspirals (EMRIs). Thus, Challenge 2 requires analysts to tackle the \emph{global analysis} of many simultaneous and superimposed signals, in the presence of the most complex sources (EMRIs) that we expect to observe with LISA.

\begin{figure}[htb]
\centerline{\includegraphics[width=10cm]{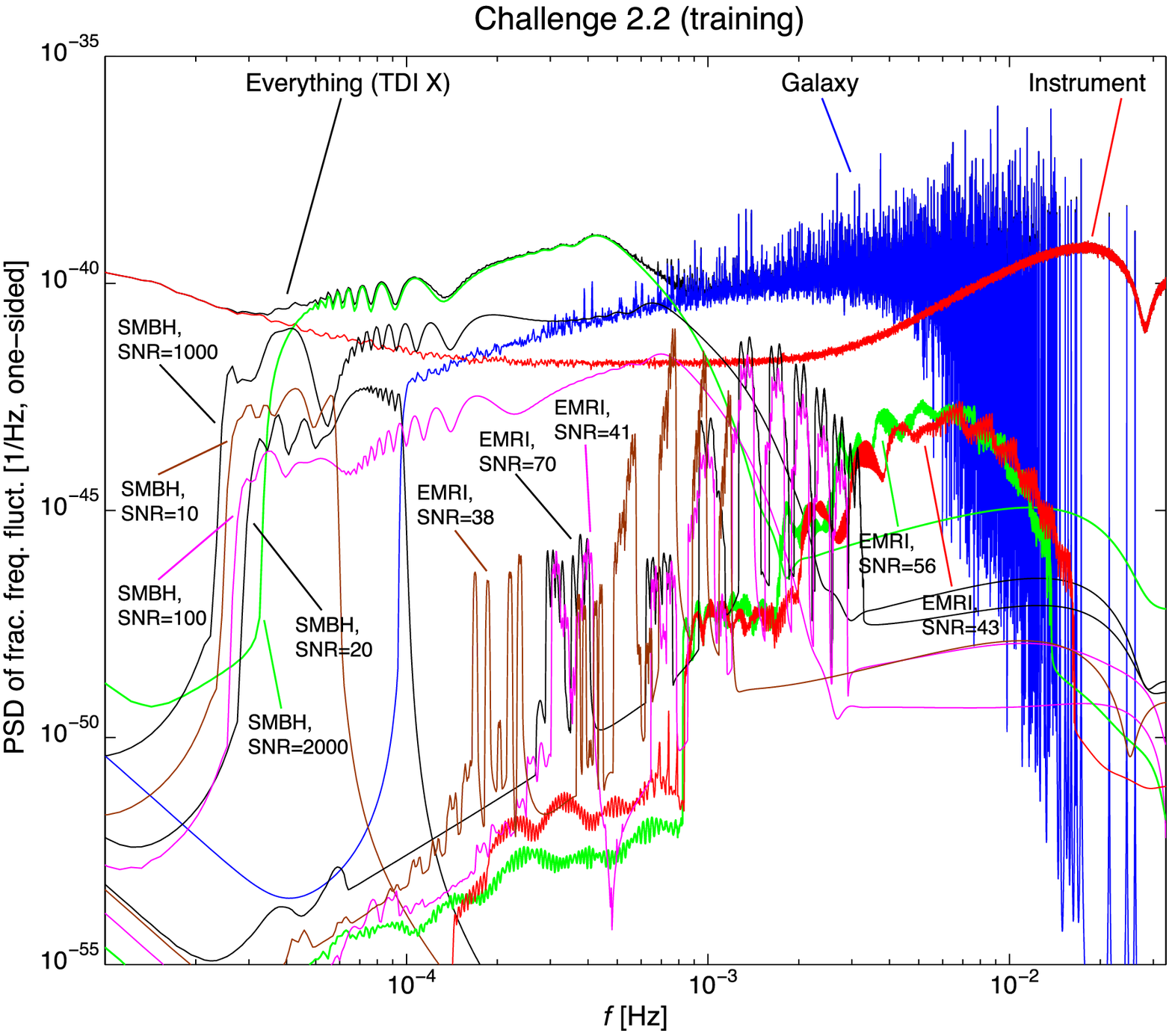}}
\caption{\protect\footnotesize
A representative Challenge-2.2 data set, plotted as the fractional frequency fluctuation spectrum of the full TDI $X$ signal, and separately of the contributions from instrument noise and from the individual sources (bundling all Galactic binary sources as the ``Galaxy'').}
\label{f:C2.2}
\end{figure}

Challenge 2 includes five single-source data sets (Challenge 1.3),\footnote{These data sets are still identified as 1.3.X, since training data sets for EMRIs were distributed with Challenge 1; blind data sets are however being released only now. For consistency with previous documents, we maintain the original nomenclature.} dedicated specifically to EMRIs, and two multi-source data sets (Challenge 2.1 and 2.2). The data sets are all $\approx 2$ years long: namely, they contain $2^{22}$ data points sampled at a cadence of 15 s. More specifically, the data sets are as follows: 
\begin{itemize}
\item \emph{Challenge 1.3} consists of five data sets, each containing a GW signal from a single EMRI with signal-to-noise ratio (SNR)\footnote{We compute SNRs for single unequal-Michelson TDI variables, selecting for each source the variable ($X$, $Y$, or $Z$) that yields the highest SNR.} between 40 and 110;
\item \emph{Challenge 2.1} contains signals from (i) a full population of Galactic binary systems (about 26 million sources), including (ii) 25 ``verification binaries'';
\item \emph{Challenge 2.2} contains signals from (i) a full population of Galactic binary systems (about 26 million sources), including (ii) 25 ``verification binaries''; (iii) an undisclosed number (between 4 and 6) of MBH binaries with SNRs between $\approx 10$ and $\approx 2000$ and different coalescence times (not all within the two years of observation); and (iv) 5 EMRIs, with SNRs between 30 and 100.
\end{itemize}
Details about the ranges from which the source parameters were drawn randomly are given in table~\ref{t:MLDC2}. Figure \ref{f:C2.2} shows a representative Challenge-2.2 data set (not used in the actual challenge). Since data sets 2.1 and 2.2 contain signals from millions of Galactic binaries, the Task Force has identified four restricted frequency bands on which analysis should concentrate first, and over which the evaluation of responses will be carried out in much more depth: these windows  are 0.2985 mHz $\le f \le$ 0.3015 mHz, 0.9985 mHz $\le f \le$ 1.0015 mHz, 2.9985 mHz $\le f \le$ 3.0015 mHz, and 5.9985 mHz $\le f \le$ 6.0015 mHz.  
\begin{table}
\caption{Summary of data sets and source-parameter ranges in Challenge 2. All angular parameters are drawn randomly from uniform distributions over the whole relevant range. Source distances are set to provide the SNR at which the signals were designed to appear. In this table, the time of coalescence $t_c$ is given relative to the time at the beginning of the observation (i.e., the time stamp of the first data sample); $U[\cdot,\cdot]$ stands for uniform distribution within the given range. Notice that EMRIs drawn from the same parameter ranges in data sets 1.3.X and 2.2 differ in SNR.}
\label{t:MLDC2}
\lineup
\flushright
\begin{tabular}{llll}
\br
Dataset & Sources & Parameters \\
\mr
\textbf{1.3} & \textit{EMRIs} & $\mu/M_\odot \in U[9.5,10.5]$, $S/M^2 \in U[0.5, 0.7]$ \\
&                                             & time at plunge $\in  U[2^{21},2^{22}] \times 15$ s \\
&                                             & eccentricity at plunge $\in U[0.15, 0.25]$ \\
1.3.1 &\multicolumn{1}{r}{\ldots one source with}         & $M / 10^7 M_\odot \in U[0.95,1.05]$, SNR $\in U[40,110]$ \\
1.3.2 &\multicolumn{1}{r}{\ldots one source with}         & $M / 10^6 M_\odot \in U[4.75,5.25]$, SNR $\in U[70,110]$ \\
1.3.3 &\multicolumn{1}{r}{\ldots one source with}         & $M / 10^6 M_\odot \in U[4.75,5.25]$, SNR $\in U[40,60]$ \\
1.3.4 &\multicolumn{1}{r}{\ldots one source with}         & $M / 10^6 M_\odot \in U[0.95,1.05]$, SNR $\in U[70,110]$ \\
1.3.5 &\multicolumn{1}{r}{\ldots one source with}         & $M / 10^6 M_\odot \in U[0.95,1.05]$, SNR $\in U[40,60]$ \\
\mr
\textbf{2.1}
& \textit{Galactic binaries} & drawn from population (see Sec~\ref{ss:galaxy})\\
&\multicolumn{1}{r}{$\sim 3\times 10^7$ sources} & \\[3pt]
& \textit{Verification binaries} & parameters in XML file posted on~\cite{MLDCweb}\\
&\multicolumn{1}{r}{25 sources} \\
\mr
\textbf{2.2}
& \textit{Galactic binaries} & drawn from population (see Sec~\ref{ss:galaxy})\\
&\multicolumn{1}{r}{$\sim 3\times 10^7$ sources} \\[3pt]
& \textit{Verification binaries} & parameters in XML file posted on~\cite{MLDCweb} \\
&\multicolumn{1}{r}{25 sources} & \\[3pt]
& \textit{MBH binaries} & $m_1/10^6\,M_\odot \in U[1,5]$, $m_2/m_1 \in U[1,4]$ \\
&\multicolumn{1}{r}{\ldots source n.\ 1} &  $t_c \in U[60,120]$ days, SNR $\sim 2000$ \\
&\multicolumn{1}{r}{\ldots source n.\ 2} &  $t_c \in U[750,780]$ days, SNR $\sim 20$ \\
&\multicolumn{1}{r}{\ldots and 2--4 out of these 4:} & \\
&\multicolumn{1}{r}{\ldots source n.\ 3} &  $t_c \in U[180,720]$ days, SNR $\sim 1000$ \\
&\multicolumn{1}{r}{\ldots source n.\ 4} &  $t_c \in U[180,720]$ days, SNR $\sim 200$ \\
&\multicolumn{1}{r}{\ldots source n.\ 5} &  $t_c \in U[495,585]$ days, SNR $\sim 100$\\
&\multicolumn{1}{r}{\ldots source n.\ 6} &  $t_c \in U[810,840]$ days, SNR $\sim 10$ \\[3pt]
& \textit{EMRIs} & $\mu/M_\odot \in U[9.5,10.5]$, $S/M^2 \in U[0.5, 0.7]$ \\
&                                             & time at plunge $\in U[2^{21},2^{22}] \times 15$ s \\
&                                             & eccentricity at plunge $\in U[0.15, 0.25]$ \\
&\multicolumn{1}{r}{\ldots one source with}          & $M / 10^7 M_\odot \in U[0.95,1.05]$, SNR $\in U[30,100]$ \\
&\multicolumn{1}{r}{\ldots two sources with}         & $M / 10^6 M_\odot \in U[4.75,5.25]$, SNR $\in U[30,100]$ \\
&\multicolumn{1}{r}{\ldots two sources with}         & $M / 10^6 M_\odot \in U[0.95,1.05]$, SNR $\in U[30,100]$ \\
\br
\end{tabular}
\end{table}

As in Challenge 1, we release both blind challenge data sets (with undisclosed source parameters) and training data sets (with public parameters selected randomly within the same ranges). The training data sets come in two flavors: ``noisy'' and noise-free, the latter containing exactly the same GW signal(s) as those present in the noisy set. In addition, to facilitate the development and testing of analysis schemes, training data sets 2.1 and 2.2 will contain the same realization of the Galaxy. This will not be the true for the blind sets. 

For each challenge data set the three TDI channels $X$, $Y$ and $Z$ are distributed through the MLDC website~\cite{MLDCweb}. The MLDC data sets are encoded in a format implemented using XML (the eXtensible Markup Language), a simple, flexible and widely used text format related to HTML \cite{xml}. Extensive software libraries to handle XML are readily available. The XML implementation of the MLDC file format (known as lisaXML) is based on XSIL (the eXtensible Scientific Interchange Language) \cite{xsil}. A number of dedicated software tools to read and write lisaXML files (for C/C++, Python, MATLAB, and for conversion to ASCII) were developed by the MLDC Task Force and are available in the LISAtools Subversion archive~\cite{lisatools}. The archive also includes all the software used for the data production pipelines (see the Task Force wiki \cite{MLDCwiki} for its usage), including the waveform-generation codes described in the next sections, as well as a variety of other software tools useful to analyze MLDC data sets.

The actual generation of the data sets is the responsibility of one member of the Task Force who does not take part in the challenges. The deadline for submission of results (again through the MLDC website) is June 15th, 2007. The Task Force plans to process the results and provide an initial summary and evaluation of this round within about a month of receiving the results.

\section{Modeling of LISA: Pseudo-LISA}
\label{s:lisa}

We have developed a set of conventions to describe the LISA orbit and response, which constitute the \emph{pseudo-LISA} adopted in Challenges 1 and 2. The pseudo-LISA orbits are obtained by truncating exact Keplerian orbits for a point mass orbiting the Sun to first order in the eccentricity (see the Appendix of~\cite{lisasimulator}). The two simulators used for the generation of the data sets -- the LISA Simulator \cite{lisasimulator} and Synthetic LISA \cite{synthlisa} -- comply with these assumptions, and adhere to these conventions. In Solar-System Barycentric (SSB) coordinates (with the $x$ axis aligned with the vernal point), we set
\begin{eqnarray}
x_n &=& a\cos \alpha + a \, e\left(\sin\alpha\cos\alpha\sin\beta_n
-(1+\sin^2\alpha)\cos\beta_n\right), \nonumber \\
y_n &=& a\sin \alpha + a \, e\left(\sin\alpha\cos\alpha\cos\beta_n
-(1+\cos^2\alpha)\sin\beta_n\right), \\
z_n & = & -\sqrt{3} \, a \, e \cos(\alpha-\beta_n) \, , \nonumber
\end{eqnarray}
where $\beta_n = (n-1)\times2\pi/3 + \lambda$ (with $n=1, 2, 3$) is the relative orbital phase of the $n$-th spacecraft, $a = 1$ AU is the semi-major axis of the guiding center, $\alpha(t)=2 \pi t / (1 \, \mathrm{year}) + \kappa$ is its orbital phase, and $t$ is time measured at the SSB. In this approximation, the spacecraft form a rigid equilateral triangle with side length $L = 2\sqrt{3} \, a \, e = 5\times 10^6$ km for $e=0.00965$. (In fact, the LISA Simulator and Synthetic LISA implement $e^2$-accurate orbits, but the additional terms make very little difference to the instrument response.) The LISA Simulator parameters $\kappa$ and $\lambda$ (\texttt{InitialPosition} and \texttt{InitialRotation} in lisaXML) set the initial location and orientation of the LISA constellation; in Challenges 1 and 2, $\kappa=\lambda=0$. This choice places LISA at the vernal point at time $t=0$, with spacecraft 1 directly below the guiding center in the southern ecliptic hemisphere.\footnote{The mapping to the Synthetic LISA \cite{synthlisa} parameters is $\eta_0 = \kappa$, $\xi_0 = 3\pi/2 - \kappa + \lambda$, $sw < 0$; the mapping to Cutler's 1996 model \cite{cutler98} is $\bar{\phi}_0 = \kappa$, $\alpha_0 = 3\pi/4 + \kappa - \lambda$.}
\begin{table}
\caption{Common source parameters. Note that in the initial challenges we do not deal explicitly with the redshifting of sources at cosmological distances; thus, $D$ is a \emph{luminosity} distance, and the masses and frequencies of table \ref{tab:bbh} are those measured at the SSB, which are red/blue-shifted by factors $(1+z)^{\pm 1}$ with respect to those measured locally near the sources.\label{tab:common}}
\flushright
\begin{tabular}{llll}
\br
{Parameter} &
{Symbol} &
{Standard parameter name} &
{Standard unit} \\
& & (lisaXML descr.) & (lisaXML descr.) \\
\mr
Ecliptic latitude   & $\beta$   & \texttt{EclipticLatitude}  & \texttt{Radian} \\
Ecliptic longitude  & $\lambda$ & \texttt{EclipticLongitude} & \texttt{Radian} \\
Polarization angle  & $\psi$    & \texttt{Polarization}      & \texttt{Radian} \\
Inclination         & $\iota$   & \texttt{Inclination}       & \texttt{Radian} \\
Luminosity distance & $D$       & \texttt{Distance}          & \texttt{Parsec} \\
\br
\end{tabular}
\end{table}

The one-way measurements between adjacent spacecraft that are necessary to build the LISA response to GWs are taken to be either the phase response $\Phi_{ij}$ (as employed in the LISA Simulator, see Sec.\ II of~\cite{lisasimulator}) or the fractional frequency response $y^\mathrm{gw}_{slr}$ (as used in Synthetic LISA, see Sec.\ II B of~\cite{synthlisa}). The TDI Rosetta Stone~\cite{rosetta} provides details for translations between index notations. The phase and fractional-frequency formalisms are equivalent, and are related by a simple time integration.\footnote{However, the LISA Simulator and Synthetic LISA adopt different sign conventions: $\Phi_{ij}$ is given by the local-laser phase minus the incoming-laser phase, while $y_{slr}$ by the incoming-laser frequency minus the local-laser frequency. The final GW responses do end up being consistent, since the two simulators use different signs in their definition of the GW polarization tensors [see (\ref{eq:defpol})].
The \emph{LISA Simulator} produces \emph{equivalent-strain} data, with a nominal length of $L_n = 10^{10}$ m; to convert equivalent strain to fractional frequency one needs to differentiate and multiply by $L_n / c$. The additional factor of $2 \pi$ given in \cite{MLDCLISA06b} for this conversion is incorrect.}

LISA employs Time-Delay Interferometry (TDI) to suppress the otherwise overwhelming laser phase noise (see~\cite{firstgen,modified,secondgen} and references therein). TDI observables are constructed from time-delayed linear combinations of one-way measurements, and they represent synthesized interferometers where laser phase fluctuations move in closed paths across the LISA arms. More complicated paths are required to deal with the real-orbit variations of the armlengths, giving rise to the three TDI ``generations.'' For the initial challenges, we adopt \emph{TDI 1.5} observables \cite{modified,secondgen}. In particular, the data sets are the unequal-arm Michelson observables $X$, $Y$, and $Z$ defined in~\cite{secondgen}. Strictly speaking, TDI 2.0 would be required to completely cancel laser noise in a flexing LISA array, such as that modelled by the simulators; however, the increase in complexity between TDI 1.5 and 2.0 complicates the numerical treatment of one-way measurements, but it is negligible for the purpose of the GW response. For the sake of simplicity, Challenges 1 and 2 feature data sets of TDI 1.5 observables with no laser noise.

The model of the LISA instrument noise adopted in Challenge 2 is identical to that considered for Challenge 1. It includes contributions from optical noise (assumed white in phase), with one-sided spectral density
\be
S_\mathrm{opt}^{1/2}(f) = 20 \times 10^{-12} \, \mathrm{m}\, \mathrm{Hz}^{-1/2},
\label{eq:optn}
\ee
and from acceleration noise (assumed white in acceleration, but increasing as $1/f$ below $10^{-4}$ Hz), with one-sided spectral density 
\be
S_\mathrm{acc}^{1/2}(f) = 3 \times 10^{-15} [1 + (10^{-4}\,{\rm Hz}/f)^2]^{1/2}\, \mathrm{m}\, \mathrm{s}^{-2}\, \mathrm{Hz}^{-1/2}.
\label{eq:accn}
\ee
As mentioned above, we do not model laser phase noise. The six optical noises and six acceleration noises (for the two optical benches on each spacecraft) are treated as independent Gaussian random processes with variances given by (\ref{eq:optn}) and (\ref{eq:accn}), and are realized in practice with sequences of pseudo-random numbers. Specifically, Synthetic LISA generates independent Gaussian deviates (i.e., white noise) in the time domain, and then filters them digitally to obtain the desired spectral shape; the LISA Simulator generates independent Gaussian deviates in the frequency domain, multiplies them by $S^{1/2}(f)$, and FFTs to the time domain.

\section{Gravitational waveforms}

In this section we describe the conventions adopted to describe the gravitational waveforms and the assumptions made on the signals to construct the data sets. They are identical to those adopted in Challenge 1, but for this challenge we introduce two new ingredients: the model of the Galaxy used to simulate a population of stellar-mass binaries, and the kludge EMRI waveforms.

\subsection{Gravitational-wave polarizations}
\label{ss:polarizations}

The sky location of a GW source is described in J2000 \emph{ecliptic coordinates}: to wit, the latitude $\beta$ and longitude $\lambda$, the latter measured from the vernal point, aligned with the $\hat{x}$ axis in our convention. Gravitational radiation travels along the direction $\hat{k} = -(\cos \beta \cos \lambda, \cos \beta \sin \lambda, \sin \beta)$, with surfaces of constant phase given by $\xi = t - \hat{k} \cdot x$. In the transverse--traceless gauge, the gravitational strain tensor can be decomposed in two polarization states $h_{+}(\xi)$ and $h_{\times}(\xi)$, and is given by\footnote{In the version of Synthetic LISA available at the time of releasing Challenge 2, the polarization tensors have opposite signs with respect to (\ref{eq:defpol}), and to the LISA Simulator convention. This minus sign cancels out the sign difference in the definition of the basic phase and frequency measurements.} 
\begin{equation}
\label{eq:defpol}
\mathbf{h}(\xi) = h_{+}(\xi) \left[ \hat{u}\otimes \hat{u} - \hat{v}\otimes \hat{v} \right] + h_{\times}(\xi) \left[ \hat{u}\otimes \hat{v} + \hat{v}\otimes \hat{u} \right],
\end{equation}
where $\hat{u} = \partial \hat{k} / \partial{\beta}$, $\hat{v} \propto \partial \hat{k} / \partial{\lambda}$. Thus, GWs from any MLDC source are completely specified by $\beta$, $\lambda$, and by the two functions $h_+(\xi)$ and $h_\times(\xi)$ for the source GW polarization amplitudes, measured at the SSB.

The orbital orientation of nonprecessing binaries is described by the inclination $\iota$ (the angle between the line of sight $\hat{k}$ and the orbital angular momentum of the binary), and by their polarization angle $\psi$: specifically, if $h^S_{+}(\xi)$ and $h^S_\times(\xi)$ are the binary's GW polarizations in its source frame (i.e., the polarizations defined with respect to the binary's \emph{principal polarization axes} $\hat{p}$ and $\hat{q}$) then 
\begin{equation}
\label{eq:polrot}
h_+(\xi) + i h_\times(\xi) = e^{-2 i \psi} \left[ h^S_+(\xi) + i h^S_\times(\xi) \right],
\end{equation}
with $\psi = -\arctan(\hat{v} \cdot \hat{p} / \hat{u} \cdot \hat{p})$.
Together with $\beta$, $\lambda$, and with the luminosity distance $D$, $\iota$ and $\psi$ form a set of common standard parameters, listed in table \ref{tab:common} with their standard lisaXML descriptors (see \cite{MLDCLISA06b} for a description of the XML files adopted for the MLDCs).

\subsection{Galactic stellar mass binaries}
\label{ss:WD}

In Challenges 1 and 2, a Galactic stellar mass binary system with component masses $m_1$ and $m_2$ at a distance $D$  is modelled as a system of two point masses in circular orbit with constant period. The source-frame polarization amplitudes are given by
\begin{eqnarray}
h^S_+(\xi)  & = & \mathcal{A} \left(1 + \cos^2{\iota}\right) \cos(2\pi f \xi + \phi_0), \\
h^S_\times(\xi) & = & -2 \mathcal{A} (\cos{\iota}) \sin(2\pi f \xi + \phi_0), \nonumber
\end{eqnarray}
where the amplitude is derived from the physical parameters of the source as $\mathcal{A} = (2 \mu / D) (\pi M f)^{2/3}$, with $M = m_1 + m_2$ the total mass, and $\mu = m_1 m_2 / M $ the reduced mass; $ \phi_0$ is an arbitrary random initial phase. Notice that $f$ is constant in the SSB, but not in the final LISA data set, because of the Doppler shifts induced by the LISA orbital motion.
A Challenge-2 \texttt{GalacticBinary} source is completely determined by the parameters listed in tables \ref{tab:common} and \ref{tab:galactic}.
\begin{table}
\caption{\texttt{GalacticBinary} source parameters. Note that \texttt{Amplitude} effectively replaces the standard \texttt{Distance} parameter.\label{tab:galactic}}
\begin{indented}
\lineup
\item[]\begin{tabular}{llll}
\br
{Parameter} &
{Symbol} &
{Standard parameter name} &
{Standard unit} \\
& & (lisaXML descr.) & (lisaXML descr.) \\
\mr
Amplitude           & $\mathcal{A}$ & \texttt{Amplitude}    & \texttt{1} (GW strain) \\
Frequency           & $f$           & \texttt{Frequency}    & \texttt{Hertz} \\
Initial GW phase    & $\phi_0$      & \texttt{InitialPhase} & \texttt{Radian} \\
\br
\end{tabular}
\end{indented}
\end{table}

Data sets 2.1 and 2.2 contain 25 ``verification binaries,'' defined as systems whose location and orbital period are exactly known (conceptually, from electromagnetic observations); the remaining four parameters are selected randomly. We consider five real known systems and then select 20 more from a synthetic model of the Galaxy (see below). In addition to verification binaries, about $26$ million
white-dwarf binaries are included from the population-synthesis model. The same set of stellar mass binaries have been used in the training sets for Challenges 2.1 and 2.2 to allow participants to compare the source recovery with and without signals of other types being present. Different realizations of the stellar-mass binaries are used in each of the blind data sets.

\subsection{Galactic model}
\label{ss:galaxy}

The Challenge-2 model of the Galaxy is derived from a modern population-synthesis code~\cite{gijs}, and contains some 26 million white-dwarf binaries. Rather than re-running the simulation each time a new realization was needed, we used a single simulation ouput, and created
new versions by tweaking the frequencies of the binaries by a random amount (of order 3 $\mu$Hz),
and by randomly drawing new values for the inclination, polarization and initial phase. These perturbations are large enough to make the
simulated galaxies distinct from the perspective of data analysis, yet small enough to leave the population unaffected in an astrophysical sense. 

Since the LISA Simulator and Synthetic LISA take a few minutes to process a single source, it is not practical to use them to generate the response to a full galactic background. Instead we used a new simulation tool~\cite{cornishlittenberg} that was designed especially to model slowly evolving signals. The output of this new tool has been checked against the LISA Simulator and Synthetic LISA, and the results agree very accurately. However, the dedicated code is able to process the entire Galaxy in just a few hours on a single processor.

\subsection{Massive--black-hole binaries}

In Challenges 1 and 2, we restrict massive-black hole binaries to having circular orbits and nonspinning black holes, and we consider only the inspiral phase of the whole coalescence process. We model the inspiral at the restricted second post-Newtonian (2PN) order, following~\cite{Blanchet,DIS}. In terms of the rescaled time
\be
\tau =  \frac{\eta}{5M}(t_c - t),
\ee
where $t_c$ is the time at coalescence, and $\eta = \mu/M$, the time evolution of the \emph{orbital} angular frequency $\omega$ and phase $\Phi$ is given at 2PN order by
\bea
M\omega &=& \frac{1}{8} \tau^{-3/8}\Biggl\{ 1 + \left( \frac{11}{32}\eta + \frac{743}{2688}\right) 
\tau^{-1/4} - \frac{3}{10}\pi\tau^{-3/8} \nonumber \\
& & + \left(\frac{1855099}{14450688} + \frac{371}{2048}\eta^2 
+ \frac{56975}{258048}\eta\right)
\tau^{-1/2}  \Biggr\}
\label{fr}
\eea
and
\bea
\Phi &=& -\frac{1}{32\eta}(M\omega)^{-5/3}\Biggl\{ 1 + \left( \frac{3715}{1008} + \frac{55}{12}\eta\right)
(M\omega)^{2/3} - 10\pi(M\omega)  \nonumber \\
&& + \left( \frac{15293365}{1016064} + \frac{27145}{1008}\eta + \frac{3085}{144}\eta^2\right)
(M\omega)^{4/3}\Biggr\}.
\label{eq:Phi}
\eea
The phase of the polarization functions is then computed according to $\Phi(t) - \Phi(t = 0) + \Phi_0$, where $\Phi_0$ is an arbitrary initial orbital phase:
\begin{eqnarray}
h^S_{+}(\xi) &=& \frac{2\mu}{D}[M\omega(\xi)]^{2/3}(1+\cos^2 \iota)\cos [2\Phi(\xi)], \\
h^S_{\times}(\xi) &=& -\frac{2\mu}{D}[M\omega(\xi)]^{2/3}(2 \cos \iota) \sin [2\Phi(\xi)]. 
\end{eqnarray}
In order to avoid numerical artifacts at the end of the inspiral (taken to be the last stable orbit), we taper the signal according to
\begin{equation}
w(t) = \frac{1}{2}\,\left( 1 - \tanh\left[A (M/R - M/R_\mathrm{taper})\right] \right),
\end{equation}
where $R$ is approximated with Kepler's law ($R = M^{1/3} \omega^{-2/3}$), and
where the dimensionless coefficient $A = 150$ was determined empirically to produce smooth damping; $R_\mathrm{taper}$ is set to $7\, M$. The lisaXML standard parameters for Challenge-2 \texttt{BlackHoleBinary} sources are listed in tables \ref{tab:common} and \ref{tab:bbh}.
\begin{table}
\caption{\texttt{BlackHoleBinary} source parameters. Note that the tapering radius was fixed at $R = 7M$ for Challenge 2.
\label{tab:bbh}}
\lineup
\begin{tabular}{llll}
\br
{Parameter} &
{Symbol} &
{Standard parameter name} &
{Standard unit} \\
& & (lisaXML descr.) & (lisaXML descr.) \\
\mr
Mass of first BH    & $m_1$  & \texttt{Mass1}           & \texttt{SolarMass} \\
Mass of second BH   & $m_2$  & \texttt{Mass2}           & \texttt{SolarMass} \\
Time of coalescence & $t_c$  & \texttt{CoalescenceTime} & \texttt{Second} \\
Angular orb. phase & $\Phi_0$ & \texttt{InitialAngularOrbitalPhase} & \texttt{Radian} \\
\multicolumn{1}{r}{at time $t = 0$} & & & \\
Tapering radius & $R$    & \texttt{TaperApplied}    & \texttt{TotalMass} \\ 
\br
\end{tabular}
\end{table}

\subsection{Extreme--Mass-Ratio Inspirals}

Extreme--Mass-Ratio Inspirals -- compact objects (CO) orbiting a MBH -- constitute the new source class included in the second round of MLDCs. For these, we adopt the Barack--Cutler ``analytic kludge'' waveforms \cite{BC}, where orbits are instantaneously approximated as Newtonian ellipses and gravitational radiation is given by the corresponding Peters--Matthews formula \cite{pm}, but perihelion direction, orbital plane, semi-major axis and eccentricity evolve according to post-Newtonian equations. While these waveforms are not particularly accurate in the highly relativistic regime of interest for real EMRI searches, they do exhibit the main qualitative features of the true waveforms, and they are considerably simpler to generate.  They are therefore ideal to develop and test search strategies. It is expected that any strategy that works for them could be modified fairly easily to deal with the true general-relativistic waveforms, once these become available.

In general relativity, the full two-body system is described by 17 parameters. Here we assume that the spin of the CO is negligible with respect to the MBH spin and can therefore be ignored: the signal is then described by 14 parameters, which we now list. Let us consider a CO of mass $\mu$ $(\ll M)$ orbiting a MBH of mass $M$ on an orbit with semi-major axis $a$, eccentricity $e$ and \emph{orbital} frequency $\nu$. At the Newtonian order,  $\nu = (2\pi M)^{-1} (M/a)^{3/2}$, and the \emph{orbital} mean anomaly $\Phi$ (i.e., the average orbital phase with respect to the direction of pericenter) is $\Phi(t) = 2\pi\nu (t-t_0) +\Phi_0$, where  $\Phi_0$ is the mean anomaly at $t_0$. The spin $\vec S$ of the MBH is parametrized by its magnitude $S$ (so that $0 \le S/M^2 \le 1$) and by the two polar angles $\theta_K,\phi_K$ in the SSB (here ``K'' stands for Kerr). $\vec L(t)$ represents the time-varying orbital angular momentum: its direction is parametrized by the \emph{constant} angle $\lambda$ (not to be confused with the ecliptic longitude, defined in Sec~\ref{ss:polarizations}) between $\vec L$ and $\vec S$%
\footnote{In reality, radiation reaction will impose a small time variation in $\lambda$; however, this variation is known to be very small \cite{scott1} and is neglected here.}, and by an azimuthal angle $\alpha(t)$ that describes the direction of $\hat L$ around $\hat S$. The angle $\tilde\gamma(t)$ is the (intrinsic) direction of pericenter, measured with respect to $\vec L\times\vec S$. Fixing the initial time $t_0$ (as measured in the SSB), $\nu_0$, $e_0$, $\tilde\gamma_0$, and $\Phi_0$ describe, respectively, the eccentricity, the direction of the pericenter within the orbital plane, and the mean anomaly at $t_0$. More specifically, $\tilde\gamma_0$ is the angle in the plane of the orbit from $\hat L \times \hat S$ to pericenter, and $\Phi_0$ is the mean anomaly with respect to pericenter passage. We refer the reader to figure 1 of \cite{BC} for a graphic representation of the system.

It is useful to describe the GW tensor in a time-varying frame defined with respect to the unit vectors ${\hat n} = -{\hat k}$ (pointing from the origin of the SSB to the source) and ${ \hat L}(t)$. Defining the unit vectors 
${ \hat p}$ and ${ \hat q}$ by
\begin{eqnarray} \label{pq}
{ \hat p} &\equiv& ({ \hat n}\times { \hat L})/
                    |{ \hat n}\times { \hat L}|, \nonumber\\
{ \hat q} &\equiv& { \hat p} \times { \hat n},
\end{eqnarray}
the general GW strain field at the SSB can then be written as
\begin{equation}
\label{eq:hab}
\mathbf{h}(\xi) = h_{+}(\xi) \left[ \hat{p}\otimes \hat{p} - \hat{q}\otimes \hat{q} \right] + h_{\times}(\xi) \left[ \hat{p}\otimes \hat{q} + \hat{q}\otimes \hat{p} \right].
\end{equation}
Notice that the polarization tensors $[ \hat{p}\otimes \hat{p} - \hat{q}\otimes \hat{q}]$  and $[ \hat{p}\otimes \hat{q} + \hat{q}\otimes \hat{p}]$ are functions of time. 
One can decompose (\ref{eq:hab}) into harmonic contributions at integer multiples of the orbital frequency,
\be
h_{+}\equiv \sum_n A^{+}_n, \quad h_{\times}\equiv \sum_n A^{\times}_n,
\label{harmonics}
\ee
where the $A^{+,\times}_n$ are
\begin{eqnarray} \label{A}
A^{+}_n &=&-[1+({ \hat L}\cdot{ \hat n})^2]\left[
a_n\cos(2\gamma)-b_n\sin(2\gamma)\right]
+[1-({ \hat L}\cdot{ \hat n})^2]c_n, \nonumber\\
A^{\times}_n&=& 2({ \hat L}\cdot{ \hat n})\left[
b_n \cos(2\gamma)+a_n \sin(2\gamma)\right].
\end{eqnarray}
The harmonics $A^{+,\times}_n$ are expressed as a function of the coefficients
\begin{eqnarray} \label{abc}
a_n &=& - n{\cal A}\bigl[J_{n-2}(ne)-2eJ_{n-1}(ne) \nonumber \\
       && +(2/n)J_n(ne) +2eJ_{n+1}(ne)-J_{n+2}(ne)\bigr]\cos[n\Phi(t)], \\
b_n &=& - n{\cal A}(1-e^2)^{1/2}\bigl[J_{n-2}(ne)-2J_{n}(ne)
+J_{n+2}(ne)\bigr]\sin[n \Phi(t)],
\nonumber\\
c_n &=& 2{\cal A}J_n(ne)\cos[n\Phi(t)]. \nonumber
\end{eqnarray}
Here the $J_n$ are the Bessel functions of the first kind, and $\gamma$ is an azimuthal angle measuring the direction of pericenter with respect to $\hat x \equiv [-\hat n + \hat L (\hat L\cdot \hat n)]/[1-(\hat L\cdot \hat n)^2]^{1/2}$.
The angle $\gamma$ is related to $\tilde \gamma$ by
\be
\label{beta}
\gamma = \tilde\gamma + \beta,
\ee
where $\beta$ is the angle from $\hat x \propto [\hat L(\hat L \cdot \hat n) - \hat n] $ to $(\hat L \times \hat S)$, as given by
\begin{equation}\label{sinbeta}
\sin\beta = \frac{(\cos\lambda) \hat L\cdot\hat n -\hat S\cdot \hat n }
{(\sin\lambda)\sqrt{1 - (\hat L\cdot\hat n)^2}}, \quad
\cos\beta = \frac{\hat n \cdot (\hat S \times \hat L)}
{(\sin\lambda)\sqrt{1 - (\hat L\cdot\hat n)^2} }.
\end{equation}
The overall amplitude is given by
\begin{equation} \label{calA}
{\cal A}\equiv (2\pi \nu M)^{2/3}\frac{\mu}{D}.
\end{equation}
In practice, we truncate the sums in (\ref{harmonics}) at $n=4$ when $e(t) < 0.136$, otherwise at $n = 30 \, e(t)$.

The angular-momentum direction vector $\hat L$ is not constant, since $\hat L$ precesses about the MBH spin
direction $\hat S$.  Let $\theta_L(t),\phi_L(t)$ be the angles specifying the instantaneous direction of $\hat L$, and let $\hat z$ be a unit vector normal to the ecliptic. One then obtains
\be \label{alpha}
\hat L = \hat S \, \cos\lambda +
\frac{\hat z - \hat S \cos\theta_K}{\sin\theta_K} \sin\lambda \cos\alpha
+ \frac{\hat S \times \hat z}{\sin\theta_K}  \, \sin\lambda \sin\alpha,
\ee
\noindent
and the angles $\theta_L(t),\phi_L(t)$ are given in terms of $\theta_K$, $\phi_K$, $\lambda$ and $\alpha(t)$ by
\begin{eqnarray}\label{relations3}
\cos\theta_L(t) &=& \cos\theta_K \cos\lambda
    +\sin\theta_K\sin\lambda\cos\alpha(t), \nonumber\\
\sin\theta_L(t)\cos\phi_L(t) &=&
\sin\theta_K\cos\phi_K\cos\lambda +\sin\phi_K\sin\lambda\sin\alpha(t) \nonumber \\
& & -\cos\phi_K\cos\theta_K\sin\lambda\cos\alpha(t), \\
\sin\theta_L(t)\sin\phi_L(t) &=&
\sin\theta_K\sin\phi_K\cos\lambda -\cos\phi_K\sin\lambda\sin\alpha(t) \nonumber \\
& & -\sin\phi_K\cos\theta_K\sin\lambda\cos\alpha(t). \nonumber
\end{eqnarray}
To work with these expressions one can exploit the relations
\be
\label{SdotN}
{\hat S}\cdot{ \hat n} = \cos\theta_S \cos\theta_K
+ \sin\theta_S \sin\theta_K \cos(\phi_S-\phi_K),
\ee
\be\label{ScrossLdotN}
\fl \hat n \cdot (\hat S \times \hat L) = 
\sin\theta_S \sin(\phi_K-\phi_S)\sin\lambda \cos\alpha + \frac{\hat S\cdot\hat n \cos\theta_K -\cos\theta_S}{\sin\theta_K} \, \sin\lambda \sin\alpha,
\ee
and
\bea
\label{LdotN}
\fl { \hat L}\cdot{ \hat n} & = & { \hat S}\cdot{ \hat n}\cos\lambda
+ \frac{\cos\theta_S - \hat S\cdot\hat n \cos\theta_K}{\sin\theta_K}
\, \sin\lambda \cos\alpha + \frac{(\hat S \times \hat z)\cdot \hat n}
{\sin\theta_K}  \, \sin\lambda \sin\alpha, \nonumber \\
\fl & = & \cos\theta_S \cos\theta_L +
\sin\theta_S \sin\theta_L \cos(\phi_S-\phi_L).
\eea
Note that the time-variation of $\hat{S}\cdot\hat{n}$ is very small in the extreme mass-ratio case considered here, and in our kludged approximation we approximate $\hat S$ -- and hence ${\hat S}\cdot{\hat n}$ -- as strictly constant.

The evolution of $\Phi(t)$, $\nu(t)$, $\tilde\gamma(t)$, $e(t)$, and $\alpha(t)$ is given by the PN formulas
\begin{eqnarray}
\fl \frac{d\Phi}{dt} &=& 2\pi\nu, \label{Phidot} \\
\fl \frac{d\nu}{dt} &=&
\frac{96}{10\pi}\frac{\mu}{M^3}(2\pi M\nu)^{11/3}(1-e^2)^{-9/2}
\biggl\{
\biggl[1+\frac{73}{24}e^2+\frac{37}{96}e^4\biggr](1-e^2) \nonumber \\
\fl &&+ (2\pi M\nu)^{2/3}\biggl[\frac{1273}{336}-\frac{2561}{224}e^2-\frac{3885}{128}e^4
-\frac{13147}{5376}e^6 \biggr] \nonumber \\
\fl &&- (2\pi M\nu)\frac{S}{M^2}(\cos\lambda) (1-e^2)^{-1/2}\biggl[\frac{73}{12}
+ \frac{1211}{24}e^2 +\frac{3143}{96}e^4 +\frac{65}{64}e^6\biggr]
\biggr\}, \label{nudot} \\
\fl \frac{d\tilde\gamma}{dt} &=& 6\pi\nu(2\pi\nu M)^{2/3} (1-e^2)^{-1}
\Bigl[1+(2\pi\nu M)^{2/3} (1-e^2)^{-1}(26-15e^2)/4\Bigr] \nonumber \\
\fl &&-12\pi\nu(\cos\lambda) \frac{S}{M^2} (2\pi M\nu)(1-e^2)^{-3/2},
\label{Gamdot} \\
\fl \frac{de}{dt}  &=& -\frac{e}{15}\frac{\mu}{M^2} (1-e^2)^{-7/2} (2\pi M\nu)^{8/3}
\biggl[(304+121\,e^2)(1-e^2)\Bigl(1 + 12 (2\pi M\nu)^{2/3}\Bigr) \, \nonumber \\
\fl &&- \frac{1}{56}(2\pi M\nu)^{2/3}\Bigl( 133640 + 108984\, e^2 - 25211\, e^4 \Bigr)\biggr]\, \nonumber \\
\fl && + e \frac{\mu}{M^2}\frac{S}{M^2}(\cos\lambda)(2\pi M\nu)^{11/3}(1-e^2)^{-4}
\, \biggl[\frac{1364}{5} + \frac{5032}{15}e^2  + \frac{263}{10}e^4\biggr], \label{edot} \\
\fl \frac{d\alpha}{dt} &=& 4\pi\nu \frac{S}{M^2} (2\pi M\nu)(1-e^2)^{-3/2}.
\label{alphadot}
\end{eqnarray}
For a point particle in Schwarzschild, the plunge occurs at
$a_{\rm min} = M (6 + 2e)(1-e^2)^{-1}$ \cite{ckp},
so we set
\be\label{numax}
\nu_{\rm max} = (2\pi M)^{-1}[(1-e^2)/(6 + 2e)]^{3/2} \, ,
\ee
and we shut-off the waveform when $\nu$ reaches this $\nu_{\rm max}$.

Equation (\ref{eq:hab}) expresses the waveform in terms of time-varying polarization tensors, but to generate the LISA responses it is necessary to re-express it in terms of fixed polarization tensors. This is achieved through a rotation by the polarization angle
\begin{equation}\label{psiSL}
\psi = \mathrm{arctan}\left(\frac{
\cos \theta_S \sin \theta_L \cos(\phi_S - \phi_L) - \cos\theta_L \sin\theta_S
}{
\sin\theta_L \sin(\phi_S-\phi_L)
}\right).
\end{equation}

The algorithm for constructing the MLDC EMRI waveforms works as follows. We fix a fiducial frequency $\nu_0$ at $t_0 = 0$ and choose the waveform parameters $\mu,\, M,\,S/M^2,\,e_0,\,\tilde\gamma_0,\,\Phi_0,\,\cos\theta_S,\,\phi_S,\,\cos\lambda,\,\alpha_0,\cos\theta_K,\,\phi_K$ and $D$. We then (i) solve the ODEs (\ref{Phidot})--(\ref{alphadot}) for $\Phi(t)$, $\nu(t)$, $\tilde\gamma(t)$, $e(t)$, $\alpha(t)$; (ii) use $e(t)$ and $\nu(t)$ to calculate $a_n(t), b_n(t), c_n(t)$ in (\ref{abc}); (iii) calculate $\theta_L(t),\phi_L(t)$ using (\ref{relations3}), and then obtain $\gamma(t)$ from $\tilde\gamma(t)$ using (\ref{beta})--(\ref{sinbeta}); (iv) compute the amplitude coefficients $A_n^{+,\times}$ using (\ref{A}) and (\ref{LdotN}); (v) calculate $\psi$ using (\ref{psiSL}); (vi) finally, compute $h_{+,\times}(t)$ from\footnote{The data sets initially distributed for Challenge 2 reflect an additional polarization rotation (\ref{eq:polrot}) by a randomly chosen angle. It was then realized that this polarization angle is effectively degenerate with $\theta_K$, $\phi_K$ and $\alpha_0$, so it is not possible to determine it from the analysis of data. The dataset key files (but not the TDI data) were then modified with values of $\theta_K$, $\phi_K$ and $\alpha_0$ that would have produced the \emph{same} gravitational waveforms without the additional polarization rotation.}
\begin{eqnarray}\label{final}
h_+(t) = A^+(t) \cos 2\psi(t) + A^{\times}(t) \sin 2\psi(t), \nonumber \\
h_{\times}(t) = - A^+(t) \sin 2\psi(t) + A^{\times}(t) \cos 2\psi(t).
\end{eqnarray}

The lisaXML standard parameters for Challenge-2 \texttt{ExtremeMassRatioInspiral} sources are listed in tables \ref{tab:common} and \ref{tab:emri}.
\begin{table}
\caption{\texttt{ExtremeMassRatioInspiral} source parameters. Note that EMRIs do not use the nonprecessing-binary inclination $\iota$ or polarization angle $\psi$.
Be aware of the collision between the symbols for the EMRI compact-object mass ($\mu$) and opening angle ($\lambda$), the binary reduced mass (again $\mu$), and the ecliptic longitude (again $\lambda$). Note that $\nu$ is called ``Azimuthal orbital frequency'' in the XML file and it is referred to as a radial orbital frequency in \cite{BC}.
If we reduce the equations to nonspinning MBH and circular orbits, then 
the orbital frequency corresponds to $\nu + \dot{\tilde{\gamma}}/{2\pi}$.
\label{tab:emri}}
\lineup
\begin{tabular}{llll}
\br
{Parameter} &
{Symbol} &
{Standard param.\ name} &
{Standard unit} \\
& & (lisaXML descr.) & (lisaXML descr.) \\
\mr
Mass of central BH     & $M$      & \texttt{MassOfSMBH}           & \texttt{SolarMass} \\
Mass of compact object & $\mu$  & \texttt{MassOfCompactObject}  & \texttt{SolarMass} \\
Central-BH spin        & $|S|/M^2$  & \texttt{Spin}      & \texttt{MassSquare} \\
Central-BH spin orient.\ & $\theta_K, \phi_K$ & \texttt{PolarAngleOfSpin}, & \texttt{Radian} \\
 \multicolumn{1}{r}{w.r.t. SSB frame}  &                         & \texttt{AzimuthalAngleOfSpin} & \\ \hline
& & \texttt{InitialAzimuthal}\ldots & \\                   
Radial orb.\ freq.\ at $t = 0$ & $\nu_0$ & \multicolumn{1}{r}{\ldots\texttt{OrbitalFrequency}} & \texttt{Hertz} \\
Orb.\ mean anom.\ at $t = 0$ & $\Phi_0$ & \multicolumn{1}{r}{\ldots\texttt{OrbitalPhase}} & \texttt{Radian} \\ \hline
Eccentricity at $t = 0$ & $e_0$ & \texttt{InitialEccentricity} & \texttt{1} \\
Dir.\ of pericenter at $t = 0$ & $\tilde{\gamma}_0$ & \texttt{InitialTildeGamma} & \texttt{Radian} \\
Azimuthal angle of orb.\ & $\alpha_0$ & \texttt{InitialAlphaAngle} & \texttt{Radian} \\
\multicolumn{1}{r}{ang.\ momentum at $t = 0$} & & & \\
Angle between spin  & $\lambda$ & \texttt{LambdaAngle} & \texttt{Radian} \\ 
\multicolumn{1}{r}{and ang.\ momentum} & & &\\
\br
\end{tabular}
\end{table}

\section{Conclusions}

The Mock Data Challenges (MLDCs) are aimed at stimulating the development and demonstrating the technical readiness of LISA data-analysis capabilities. The first round of MLDCs has just been completed and the second round data sets have been released. The latter provide a fairly realistic testbed for the data-analysis challenges that are novel to LISA (as opposed to ground-based detectors): in particular, the problem of detecting EMRIs and the global aspect of analyzing a multitude of simultaneous, overlapping signals.

\section*{Acknowledgments}
MB acknowledges funding from NASA Grant NNG04GD52G and support by the NASA Center for Gravitational Wave Astronomy at University of Texas at Brownsville (NAG5-13396).
NC and TL acknowledge funding from NASA Grant NNG05GI69G.
CC's and MV's work was performed at the Jet Propulsion Laboratory, California Institute of Technology, where it was performed under contract with the National Aeronautics and Space Administration. MV was supported by the LISA Mission Science Office and by the Human Resources Development Fund at JPL.

\end{document}